# Anisotropic thermo-mechanical response of layered hexagonal boron nitride and black phosphorus: application as a simultaneous pressure and temperature sensor

Hermann Muhammad[1], Mohamed Mezouar[1*], Gaston Garbarino[1], Tomasz Poreba[1], Giorgia Confalonieri[1], Matteo Ceppatelli[2,3], Manuel Serrano-Ruiz[3], Maurizio Peruzzini[3], Frédéric Datchi[4]

[1]*European Synchrotron Radiation Facility (ESRF), 71, Avenue des Martyrs, Grenoble, France*
[2]*LENS, European Laboratory for Non-linear Spectroscopy, Via N. Carrara 1, I-50019 Sesto Fiorentino, Firenze, Italy*
[3]*ICCOM-CNR, Institute of Chemistry of OrganoMetallic Compounds, National Research Council of Italy Via Madonna del Piano 10, I-50019 Sesto Fiorentino, Firenze, Italy*
[4]*Institut de Minéralogie, de Physique des Milieux Condensés et de Cosmochimie (IMPMC), Sorbonne Université, CNRS UMR 7590, MNHN, 4 place Jussieu, F-75005 Paris, France*

*Corresponding author

Abstract

Hexagonal boron nitride (hBN) and black phosphorus (bP) are crystalline materials that can be seen as ordered stackings of two-dimensional layers, which lead to outstanding anisotropic physical properties. The knowledge of the thermal equations of state of hBN and bP is of great interest in the field of 2D materials for a better understanding of the anisotropic thermo-mechanical properties and exfoliation mechanism of these materials. Despite several theoretical and experimental studies, important uncertainties remain in the determination of the thermoelastic parameters of hBN and bP. Here, we report accurate thermal expansion and compressibility measurements along the individual crystallographic axes, using *in situ* high-temperature and high-pressure high-resolution synchrotron X-ray diffraction. In particular, we have quantitatively determined the subtle variations of the in-plane thermo-mechanical parameters by subjecting these materials to hydrostatic pressure conditions and collecting a large number of data points in small pressure and temperature increments. Based on the anisotropic behavior of bP, we propose the use of this material as sensor for the simultaneous determination of pressure and temperature in the range 0-5 GPa, 298-1700 K.

## I. Introduction

Thermal equations of state (EOS) are fundamental properties of condensed matter [1] which are of great relevance in a variety of research fields including solid state physics, materials science and chemistry. They are intimately linked to the materials atomic arrangement and provide invaluable information about the nature and amplitude of the microscopic interactions. As such, thermal EOS can be directly related to first-principles calculations based on density functional theory (DFT) [2,3]. In this regard, materials with strong spatial anisotropy are of particular interest because they often exhibit exceptional directional physicochemical properties. Among them, hexagonal boron nitride (hBN) and black phosphorus (bP) have attracted great attention from chemists, physicists and materials scientists since their discovery. HBN is a layered crystalline solid isostructural to graphite. It is formed of single-atom thick layers of alternating boron and nitrogen arranged in the same hexagonal lattice (Figure 1 ab), while graphite is made of single-atom thick layers of carbon that give rise to graphene [4]. As such, these materials exhibit similar anisotropic properties. The hBN sheets present an AA' stacking pattern along the crystallographic *c*-axis, resulting in a crystalline structure with space group *P6$_3$/mmc* [5]. Despite being an electrical insulator with ~ 5.9 eV band gap, hBN is a good thermal conductor, which makes it one of the rare materials combining these two generally antagonistic physical properties. As hBN is commonly depicted as the insulating isostructure of graphene, it constitutes a material of choice in industrial applications such as coating [6] and dielectric material [7], and in some device applications [8]. HBN is also known for its chemical inertness [9-10], and tribological properties [11-12] that makes it interesting as solid lubricant [13-15] and high-temperature ceramic material [16].

Black phosphorus (bP) is the most stable allotrope of phosphorus at ambient conditions of pressure and temperature [17-18]. It has been first synthesized by subjecting white phosphorus to moderate pressure and temperature conditions (P>1.2 GPa and T>500 K) [19] and, since 2014 it has acquired increasing importance as the bulk material for the preparation of phosphorene by exfoliation techniques [20-21]. Indeed, bP exhibits a crystalline layered structure (A17 orthorhombic structure with space group *Cmce*) made by the ordered stacks of 2D puckered layers of P atoms, called phosphorene by analogy to graphene (Fig. 1 cd). As hBN, and due to its puckered structure, few layers phosphorene exhibit an outstanding anisotropic mechanical behavior such as a negative Poisson's ratio [22] and very contrasting directional Youngs's modulus [23]. In contrast to graphene [24], which needs doping to become semiconductor, and similarly to transition metal dichalcogenide (TMD) monolayers [25], such as $MoS_2$, phosphorene is a 2D semiconductor with a direct band gap value of 2 eV significantly larger than bP (0.3 eV) [20,26-27]. In addition, it presents a high carrier mobility (~1000 $cm^2V^{−1}s^{−1}$), good current on/off ratio (~$10^4$–$10^5$) and anisotropic in-plane properties, which makes phosphorene and phosphorene-based heterostructures very promising for future applications in nanophotonics, nanoelectronics [21,28-30], energy storage [31], sensing [32-33] and catalysis [34].

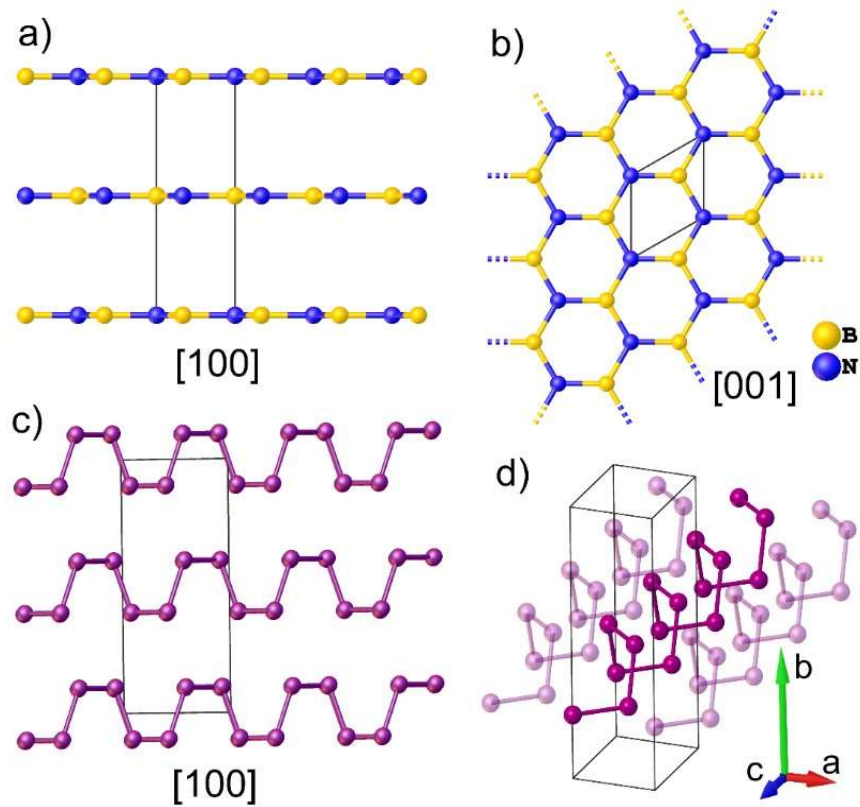

*Fig.* 1 *Crystalline structures of hBN (a,b) and bP (c,d). HBN assumes planar conformation of the layers (a) while bP forms puckered layers (c). Intra-layer differences between the regular, hexagonal layer in hBN (b) and armchair structure of bP extending in [001] (d), respectively.*

HBN and bP exhibit strong covalent intralayer bonds with $sp^2$ (hBN) and $sp^3$ (bP) hybridization and relatively weak van der Waals interlayer forces which leads to very contrasting directional thermal equations of state (EOS). The precise determination of the EOS parameters of these materials is made very difficult due to the weak pressure and temperature variations of the intra-plane inter-atomic distances. Indeed, for both materials, the intra-plane compressibility is 1 to 2 orders of magnitude lower than the inter-plane ones [35-41]. Additionally, nonlinear pressure and temperature effects have also been reported [35-52], although not yet precisely quantified.

Here we report accurate directional thermal EOS of hBN and bP using *in situ* high-resolution synchrotron x-ray diffraction performed under perfect hydrostatic conditions using liquid He as pressure transmitting medium. Besides providing an accurate description of the effects of pressure and temperature on the lattice parameters and volume of these two materials, our findings allowed to highlight a new pressure-temperature metrology based on the exceptional anisotropic properties of bP, which can be simultaneously used as pressure and temperature sensor during XRD experiments in an extended P-T domain.

## II. Experimental Method

High purity powders of hexagonal boron nitride (hBN) purchased from the company Goodfellow (product code: B-60-RD-000105), and black phosphorus (bP, 99.999+% purity) synthesized at ICCOM-CNR (Florence, Italy) from red phosphorus as described in reference [53] have been used as starting materials.

Three independent experimental runs (run 1 to 3) were carried out to determine precisely the thermal equations of state (EOS) of hBN and bP using *in situ* monochromatic X-ray diffraction. Run 1 and 2 were devoted to the determination of the ambient pressure thermal expansion coefficients of hBN and bP, while run 3 was focused on the low pressure (P<10 GPa) room temperature EOS of these materials. All the experimental work has been conducted at the European Synchrotron Radiation Facility (ESRF, Grenoble, France). Run 1 and 3 have been performed at the high-pressure XRD beamline ID27 [54] while run 2 was carried out at the high-resolution powder XRD beamline ID22 [55-56]. In the following, more details about the different runs are provided.

  a. Run 1:

This high-temperature run was carried-out by increasing the temperature at ambient pressure using the resistive heating device of a Paris-Edinburgh (PE) press [57-59]. This system can generate stable and homogeneous temperatures over a wide range from 298 to 2500 K. A standard PE sample geometry was employed [59-60]. bP was confined inside a diamond capsule of 1 mm inner diameter, 1.5 mm outer diameter and 1 mm height to ensure good thermal conductivity and minimize the temperature gradients. This diamond capsule was inserted into a chemically inert hBN cylinder, and closed using two hBN caps. No evidence of chemical reactivity of bP was observed in any of the collected x-ray diffraction patterns. The sample container was placed inside a cylindrical graphite heater of 1.9 mm inner diameter, 2.4 mm outer diameter and 3.1 mm height that was sealed using two graphite caps in contact with 2 copper electrodes. To thermally insulate the graphite heater, it was contained in an x-ray transparent boron-epoxy gasket [60]. The temperature was measured with a precision of ± 1 K using a chromel-alumel thermocouple in contact with the diamond capsule. Monochromatic x-ray diffraction measurements were performed at the beamline ID27. The incident x-ray beam energy was fixed to 20 keV ($\lambda$=0.6199 Å), covering a large scattering angle from 3 to 25°. The two-dimensional XRD patterns were collected in transmission geometry using a MAR165 CCD detector. A high efficiency multichannel collimator [61] was used to remove most of the parasitic elastic and inelastic x-ray signal coming from the sample assembly (graphite heater and boron-epoxy gasket). The sample to detector distance, detector tilt angles and beam center were accurately determined using a $LaB_6$ powder standard. The two-dimensional XRD images were integrated using the PyFAI software [62] as implemented in the DIOPTAS [63] suite. The resulting one-dimensional XRD patterns have been analyzed using the GSAS software [64]. The unit-cell parameters and volume of hBN and bP have been obtained by Le Bail [65] extraction of d-spacings using a pseudo-Voigt peak shape function. As hBN and bP were contained in the same sample assembly, the XRD patterns contain contributions from both samples. A typical XRD pattern of hBN and bP obtained at ambient pressure and 300 K, and the corresponding Le Bail adjustment are presented in Fig. 2(a). The temperature was increased from room T to 760 K by small T increments of 10 K up to 470 K, and slightly larger ones (15 K) at higher T.

b. Run 2:

In order to cross-check the results obtained in run 1 for bP, run 2 has been performed using a different experimental set up. This high-T run 2 has been carried out at the high-resolution powder diffraction beamline ID22 (ESRF) using a heating device consisting of a hot gas blower from the company Cyberstar positioned at an optimum distance (~5 mm) from the sample. This device generates a smooth laminar flow of hot air that results in a homogeneous temperature distribution over a large volume of more than 10 mm$^3$. This homogeneously heated volume is much larger than the sample (~1 mm$^3$) ensuring very small temperature gradients. BP was contained in a 0.5 mm inner diameter quartz capillary. The incident energy was set to 35 keV ($\lambda$=0.3543 Å). The temperature interval covered in this run was from 298 to 715 K as it has been previously reported that bP starts to decompose at higher T [47]. As this run was intended to validate run 1, data were obtained at wider temperature intervals of the order of 60 K (±1K). High resolution XRD patterns were collected over the 2Θ range of 1° – 62° using a multi-analyzer stage of 13 Si(111) crystals coupled to a Dectris Eiger2 CdTe pixel detector. As for run 1, the structural parameters were derived by Le Bail [65] extraction of the d-spacings using a pseudo-Voigt peak shape function using the GSAS software [64]. A typical XRD pattern obtained at 300 K and ambient pressure and the corresponding Le Bail fitting are shown in Fig. 2(b).

c. Run 3:

Run 3 was dedicated to the accurate determination of the room T equation of state (EOS) of bP and hBN. As for run 1, the x-ray diffraction measurements were conducted at the ID27 high-pressure XRD beamline. Fine powder samples of bP and hBN were loaded together in a Le Toulec type membrane diamond anvil cell [66]. The high-pressure cavity consisted of a 300 μm hole drilled in a 50 μm thick rhenium foil. Helium, which remains liquid below 11 GPa and thus provides perfect hydrostatic conditions over the whole investigated pressure range, was used as pressure transmitting medium. The pressure was determined from the shift of the $R_1$ luminescence peak of a ruby chip placed in the pressure cavity using the Ruby2020 pressure scale [67]. The pressure (P) was determined as the average of the pressures measured before and after XRD data collections with a maximum P difference of 0.1 GPa. P was increased in fine steps (0.15 GPa steps in the 0-4 GPa pressure range and 0.4 GPa steps in the 4-10.5 GPa pressure range) up to 10.5 GPa. This has been essential to accurately determine the very small in-plane cell parameters variation of hBN and bP. The incident x-ray beam energy was fixed to 33.169 keV ($\lambda$ = 0.3738 Å) using a silicon (111) double-crystal monochromator. The two-dimensional diffraction images were collected in transmission geometry using a Dectris Eiger2 9M CdTe pixel detector. The sample to detector distance, detector tilt angles and beam center were accurately determined using a CeO$_2$ powder standard. The data were analyzed in a similar way as for run 1 and 2 (Le Bail fitting).

Typical diffraction patterns of bP and hBN acquired during run 3 at room temperature and ambient pressure, respectively, and the corresponding Le Bail refinements are shown in Fig. 2(c) and Fig. 2(d), respectively.

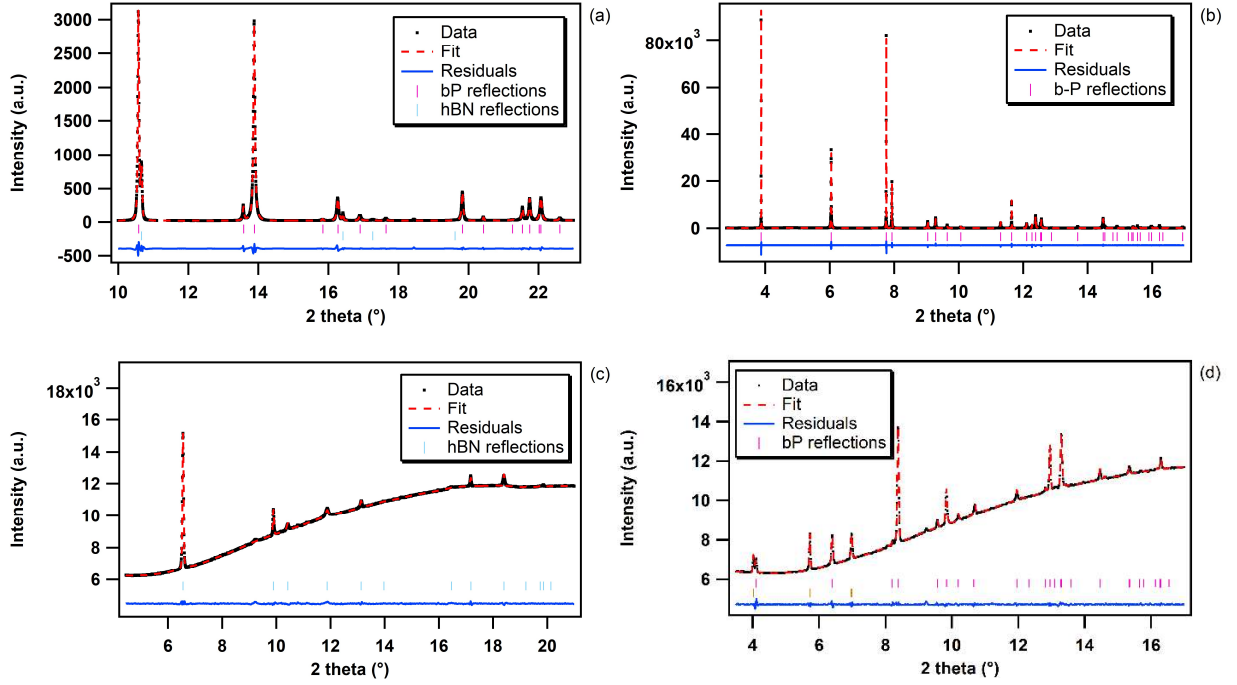

*Fig. 2 Integrated XRD patterns of hBN (panel a,c) and bP (panel b,d) in run 1 to -3 and their corresponding Le Bail fittings. (a) Run 1 (Ambient pressure, T=300 K; the XRD patterns were collected simultaneously), (b) Run 2 (Ambient pressure, T=300 K), (c) and (d) Run 3 (Ambient pressure, T=300K). The higher background in run 2 and 3 is due the small sample dimension and high x-ray Compton scattering signal from the diamond anvil cell.*

### III. Thermal expansion of hBN and bP at ambient pressure

a. Hexagonal boron nitride

The temperature dependence of the unit-cell parameters and volume of hBN obtained in run 1 is presented in Fig. 3. The resulting linear and volumetric thermal expansion coefficients are listed in Table 1.

The $c$ lattice parameter presents, within experimental error, a standard linear and positive temperature evolution up to 1280 K, the maximum temperature reached in this run (see Fig. 3b). A linear regression of the data gives:

$$c(T) = c_0[1 + \alpha_c(T - 300)] = 6.6706(7)[1 + 4.23(1) \times 10^{-5} \times (T - 300)] \quad Eq.\ 1$$

where $c_0$ is the $c$ lattice parameter at 300 K, $\alpha_c$ the corresponding linear thermal expansion coefficient, and T the temperature. Here and throughout the manuscript, the error bars refer to the standard deviation of the parameters as obtained from the mathematical refinement of the experimental data.

As previously reported [42,43], along the $a$ crystallographic axis, we observed an unusual negative variation of the thermal expansion (Fig. 3a). Indeed, along this axis the network contracts in a non-linear way, at least up to 1280 K. It is worth noting, that the absolute temperature variation of the $a$ lattice parameter is 2 orders of magnitude weaker than that of $c$. Its temperature dependence can be expressed as a third order polynomial expression:

$$a(T) = a_0[1 + \alpha_a(T - 300) + \alpha'_a(T^2 - 300^2) + \alpha''_a(T^3 - 300^3)]$$
$$= 2.5069(2)[1 - 4.63(17) \times 10^{-6} \times (T - 300) + 3.07(23) \times 10^{-9} \times (T^2 - 300^2)$$
$$-7.0(10) \times 10^{-13} \times (T^3 - 300^3)] \quad Eq.\ 2$$

where $a_0$ is the $a$ lattice parameter value at 300 K, $\alpha_a$, $\alpha'_a$ and $\alpha''_a$ the corresponding thermal expansion coefficient, its first and second derivatives respectively, and $T$ the temperature (Fig. 3a, Table 1).

Due to the weak effect along the $a$-axis, the evolution of the unit-cell volume of hBN is dominated by the temperature effect on the $c$ lattice parameter and evolves linearly with T over the probed temperature range as expressed by:

$$V(T) = V_0[1 + \alpha_V(T - 300)] = 36.290(3)[1 + 3.96(1) \times 10^{-5} \times (T - 300)] \quad Eq.\ 3$$

where $V_0$ is volume at 300 K, $\alpha_V$ the corresponding volumetric thermal expansion coefficient and T the temperature (Fig 3c, Table 1).

The large contrast between the directional thermal expansion coefficients corresponding to the $a$ and $c$ lattice parameters is explained by the anisotropic structure of hBN, particularly by the striking difference in nature between the chemical bonds in the $ab$ plane and along the $c$ axis of the hexagonal lattice. As shown in Fig. 3a and Table 1, the negative thermal expansion along the $a$-axis has already been reported [42,43] although never precisely quantified in a wide temperature domain. This is explained by the low number of collected data points and narrower temperature interval probed in these former studies. As suggested by Pease [42], this unusual variation may be due to the fact that the strong in-plane chemical bonds oppose standard thermal expansion by layer puckering effects related to out-of-plane thermal motion. The much larger and positive thermal expansion along the *c* axis is resulting from the low van der Waals forces that are acting along this crystallographic direction.

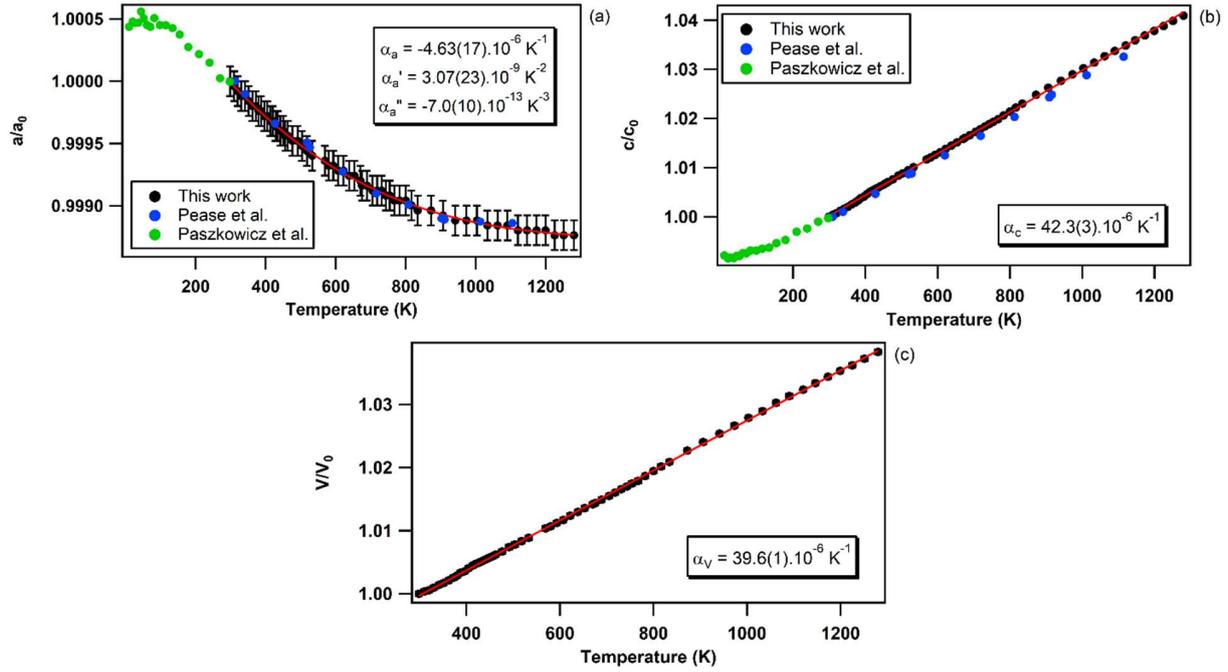

*Fig. 3* Relative thermal expansion of the a and c lattice parameters of hBN with respect to their ambient T values at ambient P, respectively along the [100] (panel a) and, [001] (panel b) directions. The volumetric thermal expansion of hBN at ambient P is displayed in panel c. Symbols correspond to the experimental data from this (black) and literature (color) studies, whereas the red curves are linear or polynomial fits of our data. Error bars in the last two panels are smaller than the size of the plotted points.

| Exp | T range (K) | $\alpha_a$ ($10^{-6}$ K$^{-1}$) | $\alpha'_a$ ($10^{-9}$ K$^{-2}$) | $\alpha''_a$ ($10^{-13}$ K$^{-3}$) | $\alpha_c$ ($10^{-6}$ K$^{-1}$) | $\alpha_V$ ($10^{-6}$ K$^{-1}$) | Technique |
|---|---|---|---|---|---|---|---|
| This work | 298-1280 | -4.63(17) | 3.07(23) | -7.0(10) | 42.3(1) | 39.6(1) | Powder XRD |
| Pease [42] | 273-1073 | -2.9 | 1.9 | - | 40.5 | - | Powder XRD |
| Zhao et al. [36] | 300-1280 | -0.0098 | 0.0102 | - | 51.6 | 49.1(19) | Powder XRD |
| Paszkowicz et al. [43] | 128-297.5 | -2.72 | 0 | - | 3.2 | - | Powder XRD |
| Solozhenko et al. [44] | 300-1800 | - | - | - | - | 40.9(8) | Powder XRD |
| Yates et al. [45] | 80-780 | -2.76 | - | - | 38.0 | - | Interferometric measurements |

*Table 1* Volumetric (V) linear (c) and non-linear (a) thermal expansion coefficients of hBN at ambient pressure from this work and the literature.

b. Black phosphorus

The relative temperature dependence of the unit-cell parameters and volume of bP obtained from Le Bail fitting [65] are presented in Fig. 4. The evolution of the linear and volumetric thermal expansions from run 1 and 2, obtained using two different experimental methods, are in excellent agreement. A linear regression is employed to determine the directional and volumetric thermal expansions in the 298-750 K temperature range. As reported in previous studies [47], the sudden reduction of the Bragg reflections intensities (not shown here), followed by their complete disappearance, attests for the decomposition of bP at higher temperature. The obtained linear and volumetric thermal expansion coefficients are listed and compared with the literature data in Table 2. The volumetric thermal expansion of bP obtained in this work is in excellent agreement with that reported by Henry et al. [47] and Rodrigues et al. [49] and in relatively good agreement with that reported by Faber et al. [48]. As for this study, the employed experimental method in [47-51] was *in situ* XRD. A good agreement is also found with the density-functional theory (DFT) calculations of Sansone et al. [52]. However, strong deviations emerge with respect to the papers by Keyes et al. [51] and Riedner et al. [50], likely due to a much smaller set of collected data points in these studies.

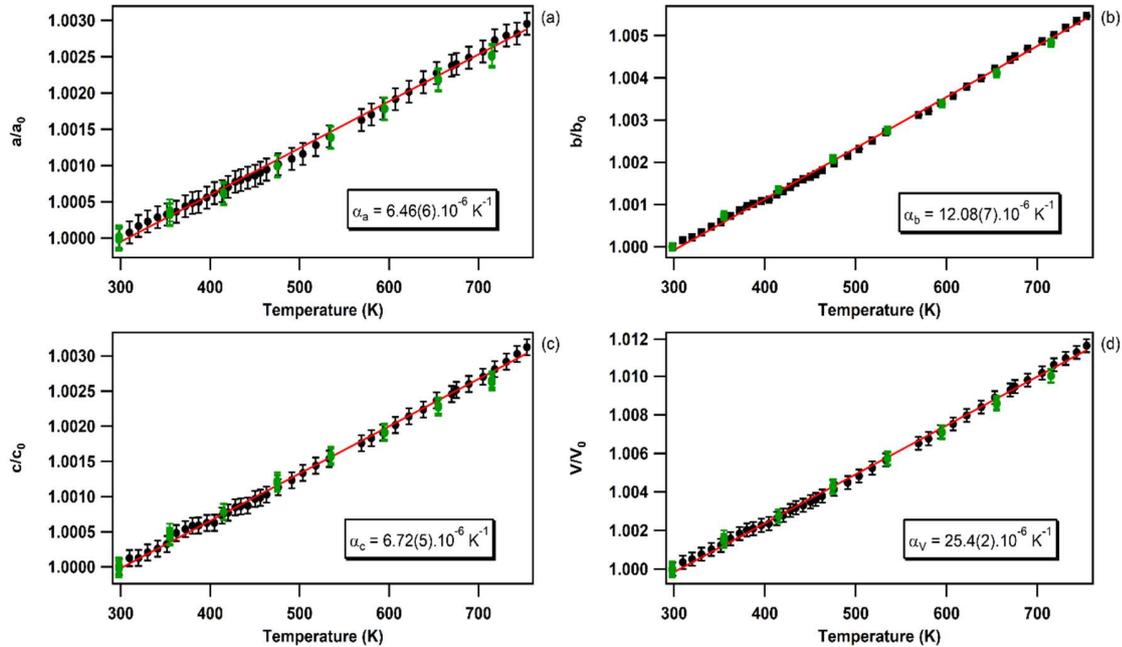

**Fig. 4** *Thermal expansion of lattice parameters (a, b and c) and volume (V) of bP at ambient pressure with respect to the corresponding ambient T values ($a_0$, $b_0$, $c_0$, $V_0$). Black and green circles correspond to data coming from run-1 and run-2 respectively.*

As expected, the smallest linear thermal expansion is found in the (010) plane, where the atoms form covalently-bonded puckered layers. By comparison, the crystallographic *b* direction, normal to the layers, has a coefficient of thermal expansion almost twice as large with respect to the *a* and *c* directions (Table 2). This is consistent with weaker van der Waals-type interactions only present along the crystallographic *b* direction. As previously reported by Henry et al. [47], we do not confirm the in-plane anisotropy suggested by the DFT calculations of Sansone et al. [52]. In contrast with Henry et al. [47], we did not observe a discontinuity in the slope of the linear expansion coefficients for the *a* and *c* parameters above 706 K. This could be due to the smaller temperature increment used in the present study which resulted in better data sampling and smaller error bars.

| Exp | T range (K) | $\alpha_a$ ($10^{-6}$ K$^{-1}$) | $\alpha_b$ ($10^{-6}$ K$^{-1}$) | $\alpha_c$ ($10^{-6}$ K$^{-1}$) | $\alpha_V$ ($10^{-6}$ K$^{-1}$) | Technique |
|---|---|---|---|---|---|---|
| This work | 298-750 | 6.46(6) | 12.08(7) | 6.72(5) | 25.4(2) | Powder XRD |
| Henry et al. [47] | 300-706 | 6.4(1) | 11.8(1) | 6.5(2) | 24.8(2) | Powder XRD |
| Faber et al. [48] | 300-575 | 8(5) | 11(2) | 5(5) | 22(12) | Powder XRD |
| Rodrigues et al. [49] | 170-250 | 4.9(3) | 11.7(4) | 7.7(2) | 24.5(3) | Powder XRD |
| Riedner et al. [50] | 300-475 | 53(5) | 10(2) | 0(5) | 63(12) | Single Crystal XRD |
| Keyes et al. [51] | 300-700 | 22(2) | 38(4) | 39(4) | 99(10) | Powder XRD |
| Sansone et al. [52] | 300-600 | 1 | 11 | 8 | 20 | DFT |

*Table 2* Volumetric and lattice thermal expansion coefficients of bP at ambient pressure from this work and the literature.

IV. **Room temperature equations of state of hBN and bP**

a. Hexagonal boron nitride

The pressure dependence of unit-cell parameters and volume of hBN from run 3 and literature are shown in Fig. 5.

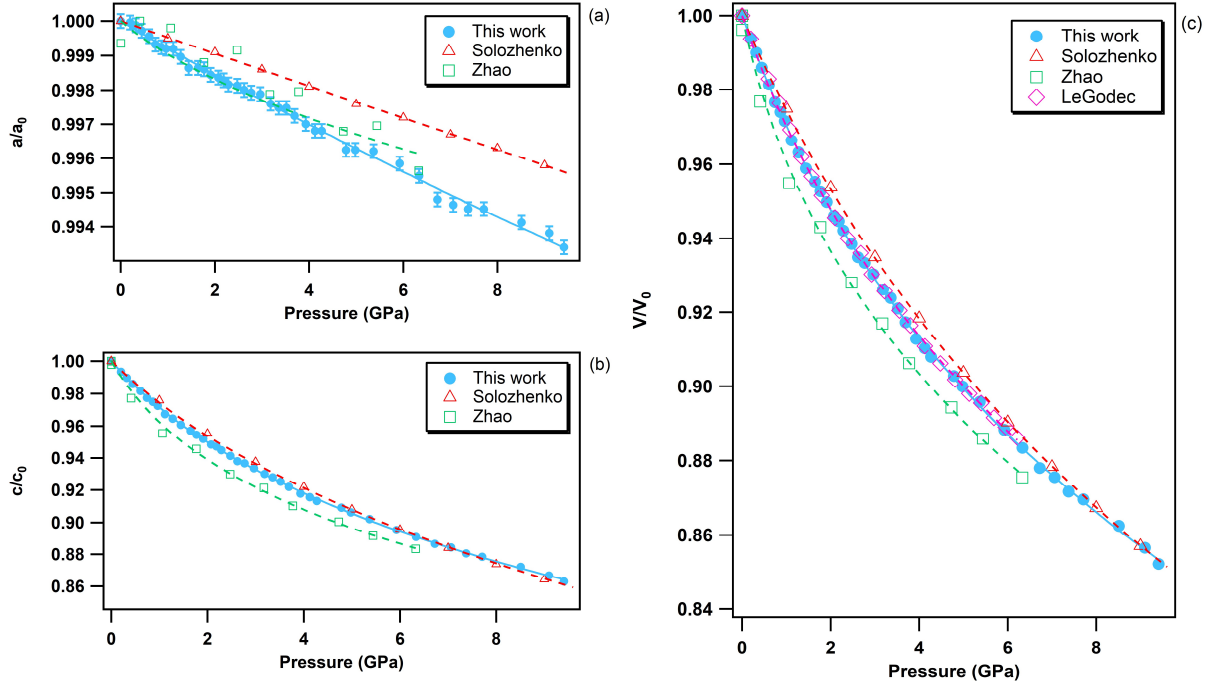

**Fig. 5** *Pressure evolution of the relative lattice parameters ($a/a_0$ and $c/c_0$) and volume of hBN in the 0-10 GPa pressure range at room temperature. The solid blue circles represent the data acquired in the present study upon compression. Empty symbols refer to the data reported in the literature. Red triangles: Solozhenko et al. [35]; green squares: Zhao et al. [36]; pink diamonds: Le Godec et al. [46]. The individual a and c-axis data were fitted using a polynomial equation and the volume data using a third order Birch-Murnaghan equation of state [68].*

As presented in Fig. 5a, the *a* lattice parameter of hBN presents a very small variation in the 0-10 GPa pressure range, only reducing by 0.7% of its initial value over the investigated pressure range. By contrast, we have measured a much larger variation of 14% along the *c* axis in the same pressure interval (Fig. 5b). The strong compressibility anisotropy can be clearly appreciated in Fig. 6a. This observation is consistent with the characteristic layered structure of the material featuring strong intra-plane covalent bonds and weak van der Waals inter-plane interactions.

| Exp | P range (GPa) | $V_0$ (Å$^3$) | $K_0$ (GPa) | $K_0'$ | Technique | EOS type |
|---|---|---|---|---|---|---|
| This work | 0 – 9.5 | 36.18 ± 0.06 | 27.4 ± 0.9 | 11.4 ± 0.8 | Powder XRD | 3$^{rd}$ order BM |
| Solozhenko et al. [35] | 0 – 12 | 36.17 | 36.7 ± 0.5 | 5.6 ± 0.2 | Powder XRD | Murnaghan |
| Zhao et al. [36] | 0 – 9.0 | - | 17.6 ± 0.8 | 19.5 ± 3.4 | Powder XRD | 3$^{rd}$ order BM |
| Le Godec et al. [46] | 0 – 6.7 | - | 27.6 ± 0.5 | 10.5 ± 0.5 | Powder XRD | 3$^{rd}$ order BM |

**Table 3** *Refined EOS parameters of hBN at ambient temperature from this work and literature.*

| Axis | P range (GPa) | A | B | C | D | Technique | Polynomial fit |
|---|---|---|---|---|---|---|---|
| a axis | 0 – 9.5 | 2.504[9] | -0.00081[2] | 0.000012[2] | - | Powder XRD | 2nd order |
| c axis | 0 – 9.5 | 6.661[5] | -0.0284[3] | 0.0025[1] | 0.000107[6] | Powder XRD | 3rd order |

*Table 4* 3rd order polynomial adjustment of the directional variation of the individual a and c-axis of hBN. A, B, C, D are the polynomial coefficients obtained from the fit.

The equation of state (EOS) parameters of hBN have been derived using the EoSFit program [69,70] by fitting the experimental data using a third-order Birch-Murnaghan EOS [68] with the following expression:

$$P = \frac{3}{2}K_0\left[\left(\frac{V}{V_0}\right)^{-7/3} - \left(\frac{V}{V_0}\right)^{-5/3}\right] \times \left\{1 + \frac{3}{4}(K_0' - 4)\left[\left(\frac{V}{V_0}\right)^{-2/3} - 1\right]\right\} \qquad Eq.\ 4$$

where $P$ is the pressure, $V_0$ and $V$ are the unit-cell volumes at ambient pressure and at pressure P, $K_0$ and $K_0'$ are the bulk modulus and its first derivative at ambient pressure, respectively. The fitted EOS parameters are listed in Table 3 together with the available literature data. Polynomial regressions of the directional unit-cell parameters are presented in Table 4. The EOS parameters obtained in this study are in very good agreement with those reported in the study of Le Godec et al. [46], in which, however, only the volumetric variations are reported, thus not allowing a direct comparison with the directional variations obtained in the present study. On the other hand, significant differences with the work of Solozhenko et al. [35] and Zhao et al. [36] are highlighted in Fig. 5. Specifically, along the *a*-axis, the scatter of data points is more pronounced in Zhao et al. [36], while a smaller slope of the linear variation is observed in Solozhenko et al. [35]. This is explained by the very small variation of the parameter *a* (0.7% over the investigated pressure range) which is at the sensitivity limit of the *in situ* XRD method employed in these studies. Moreover, in these works, a non-hydrostatic solid medium was used, which could potentially induce large pressure gradients. These in turn can lead to unprecise pressure and cell parameters determination. In the current study, helium was employed as pressure-transmitting medium, thus ensuring perfect hydrostatic conditions. Along the *c*-axis, as the variations with pressure are much larger (14%), the relative difference between our data and those of the two literature studies is less pronounced.

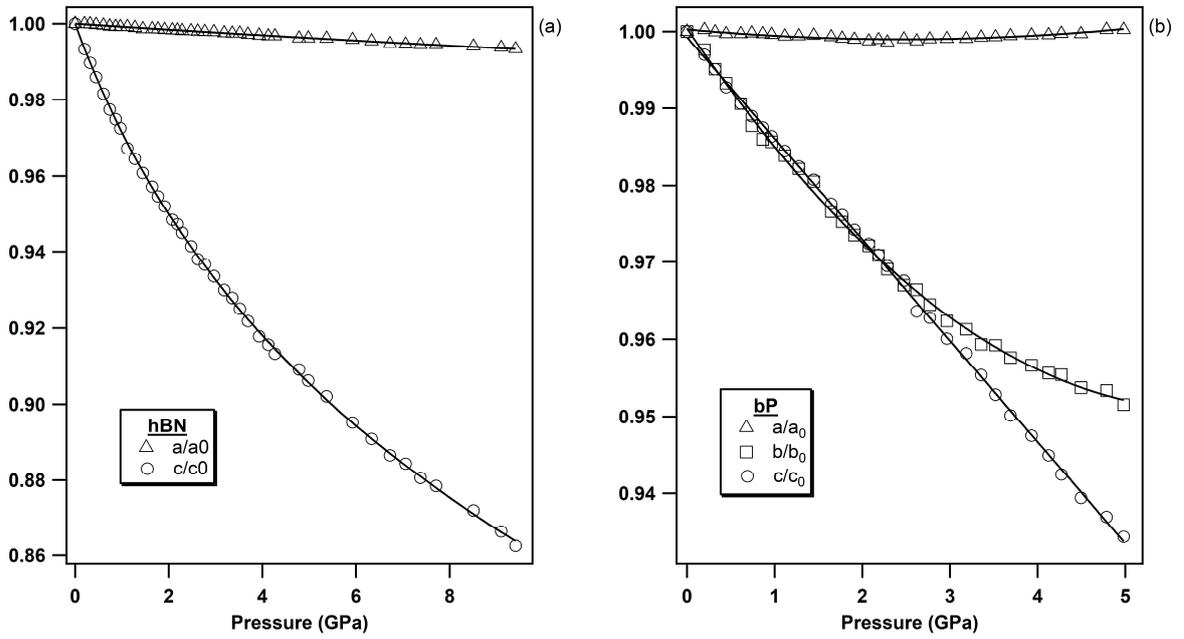

*Fig. 6* Pressure evolution of the lattice parameters of (a) hBN and (b) bP, normalized to their ambient pressure values to highlight their anisotropic compressive behavior.

b. Black phosphorus

The pressure variation of the relative lattice parameters and unit-cell volume of bP at ambient temperature are presented together with literature data in Fig. 7.

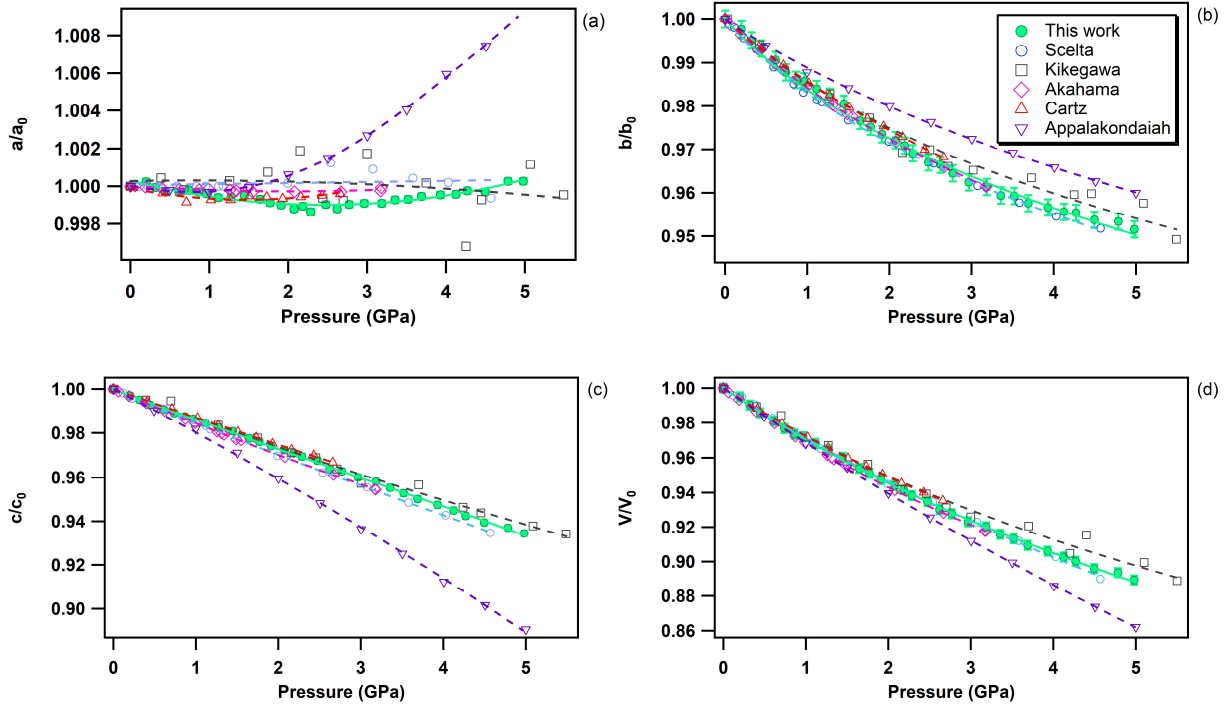

**Fig. 7** *Relative lattice parameters and unit-cell volume variation of bP in the 0-5 GPa pressure range at room temperature. The solid green circles represent the data obtained in the present study. The empty symbols refer to the data reported in the literature. Blue circles : Scelta et al. [37]; black rectangles; Kikegawa et al. [38]; pink diamonds; Akahama et al. [39]; red up triangles; Cartz et al. [40]; and, violet down triangles; Appalakondaiah et al. [41]. The individual a, b and c-axis data were fitted using a polynomial equation and the volume data using a third order Birch-Murnaghan equation of state [68].*

As hBN, bP exhibits a strong anisotropic behavior (see Fig. 6b). The *b* and *c* axes of the orthorhombic lattice reduces by approximately 5% while the *a*-axis is nearly incompressible in the entire stability field of bP up to 5 GPa. Using *ab initio* DFT calculations, Appalakondaiah et al. [41] proposed an explanation for this mechanical behavior. Indeed, they showed that the elastic constant $C_{11}$ is much larger than the $C_{22}$ and $C_{33}$ ones (see ref. [41] for $C_{ii}$ index correspondence with the crystallographic directions), which implies that bP is stiffer against deformation along the *a*-axis than along the *b* and *c* axes. Owing to the very small volume change along *a*, very accurate XRD measurements and perfect hydrostatic conditions are required for quantitative analysis. We could observe a shallow relative *a*-axis contraction of $10^{-3}$ between 0 and ~2.4 GPa, followed by an expansion of the same amount at higher pressure up to 5 GPa. As shown in Fig. 7, such behavior was not reported in previous experimental studies [37-40] due to higher data points scattering, limited pressure range, or the presence of pressure gradients. It is also worth noting that, despite a significant difference in the absolute values, the DFT calculations in ref. [41] reproduce well the observed experimental trends under pressure. This behavior may be attributed to competitive effects between the layer puckering of the sheets and of the atomic repulsion.

| Exp | P range (GPa) | $V_0$ (Å³) | $K_0$ (GPa) | $K'_0$ | Technique | EOS type |
|---|---|---|---|---|---|---|
| This work | 0 – 5 | 152.06 ± 0.16 | 29.8 ± 0.7 | 5.7 ± 0.5 | Powder XRD | 3rd order BM |
| Scelta et al. [37] | 0 – 5 | 151.28 ± 0.16 | 33.3 ± 1.3 | 3.1 ± 0.6 | Powder XRD | Vinet |
| Kikegawa et al. [38] | 0 – 5 | 151.2 ± 4.8 | 36 ± 2 | 4.5 ± 0.5 | Powder XRD | Murnaghan |
| Akahama et al. [39]* | 0 – 3.2 | 151.94 ± 0.02 | 29.1 ± 0.2 | 5.7 ± 0.2 | Powder XRD | 3rd order BM |
| Cartz et al. [40]* | 0 – 2.7 | 151.80 ± 0.17 | 32.8 ± 2.3 | 5.8 ± 1.9 | Powder XRD | 3rd order BM |
| Appalakondaiah et al. [41] | 0 – 5 | 151.3 | 30.7 | - | DFT | - |

*Table 5* Volumetric EoS parameters of bP at ambient temperature from this work and literature. *These parameters have been calculated from the data provided in the literature.

| Axis | P range (GPa) | A | B | C | D | Technique | Polynomial fit |
|---|---|---|---|---|---|---|---|
| $a$-axis | 0 – 5 | 3.314[1] | -0.0009[2] | 0.00014[8] | 0.000009[11] | Powder XRD | 3rd order |
| $b$-axis | 0 – 5 | 10.48[1] | -0.0163[6] | 0.0012[3] | 0.00003[4] | Powder XRD | 3rd order |
| $c$-axis | 0 – 5 | 4.375[2] | -0.01315[6] | - | - | Powder XRD | Linear |

*Table 6* Lattice parameters evolution of bP with pressure. A, B, C, D are the polynomial coefficients obtained in the linear (c axis) and third-order (a,b axis) adjustments.

The volumetric EOS and directional 3rd order polynomial parameters of bP are listed in Table 5 and 6, respectively. As for hBN, a third-order Birch-Murnaghan EOS [68] was employed to fit the experimental volume data. Due to the quasi-incompressible nature of the $a$-axis, the obtained polynomial coefficients B, C and D are close to zero. This unique feature can be exploited for accurate P, T metrology as detailed in section V. In agreement with previous studies [37-40], the $c$ axis has the particularity to vary quasi-linearly with pressure, while the $b$ axis presents a regular behaviour. With the exception of Kikegawa et al. [38], our volumetric EOS parameters $V_0$, $K_0$ and $K'_0$ are in good agreement with those reported in the literature.

c. A7 phase of phosphorus

BP displays a series of pressure-induced first-order phase transitions [37]. As shown in Fig. 8, upon compression orthorhombic bP undergoes a first-order phase transition to the A7 rhombohedral form (space group *R-3m*). As previously reported [37], the onset of this phase transformation is observed at ~5 GPa and is associated with a large volume discontinuity of 13% which corresponds to a major atomic rearrangement. This leads to the partial loss of the structural anisotropy of bP. The EOS parameters of the A7 phase of P are listed in Table 7 together with the literature data. Our results are in excellent agreement with those of Scelta et al. [37], but strongly deviates from those reported by Kikegawa et al. [38] and Clark et al. [71]. This is again explained by the larger data point scatter and the potential presence of significant pressure gradients as non-hydrostatic solid pressure media have been employed in these two studies. As mentioned earlier, helium was used as pressure transmitting medium both in this study and that of Scelta et al. [37].

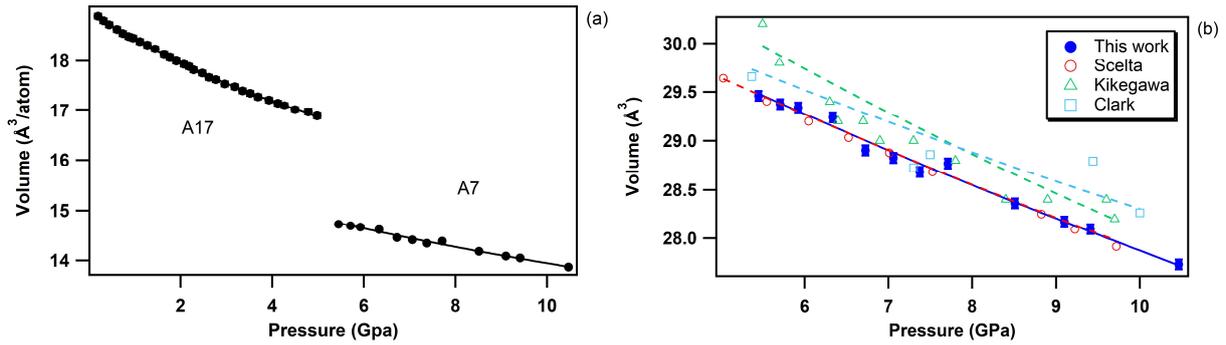

*Fig. 8* (left) Equation of state of bP at ambient temperature across the orthorhombic to rhombohedral (A7) first-order phase transition, observed at 5 GPa. (right) Unit-cell volume evolution of A7 rhombohedral P in the 5 – 10.5 GPa pressure range at room temperature. The solid blue circles represent the data acquired in the present study. The empty symbols refer to the data reported by Scelta et al. [37] (red circles), Kikegawa et al. [38] (green up triangles) and Clark et al. [71] (blue squares). The corresponding EoS fits are represented by lines.

| Exp | P range (GPa) | $V_0$ (Å³/atom) | $K_0$ (GPa) | $K_0'$ | Technique |
|---|---|---|---|---|---|
| This work | 5 – 10.5 | 15.92 ± 0.06 | 64.8 ± 2.8 | 2.4 ± 0.4 | Powder XRD |
| Scelta et al. [37] | 5 – 10 | 15.88 ± 0.02 | 68 ± 2 | 1.9 ± 0.3 | Powder XRD |
| Kikegawa et al. [38] | 5.5 – 9.7 | 16.6 ± 0.2 | 46 ± 4 | 3.0 ± 0.6 | Powder XRD |
| Clark et al. [71] | 3 – 10 | 15.97 ± 0.02 | 65.0 ± 0.6 | - | Powder XRD |

*Table 7* Volumetric EOS parameters of rhombohedral A7 phosphorus at ambient temperature from this work and literature.

# V. Exploiting the anisotropic thermo-elastic properties of bP for simultaneous P,T determination

As shown by Crichton and Mezouar [72], it is possible to simultaneously determine the pressure and temperature from the thermal equations of state of 2 materials subjected to the same P,T conditions. This method is commonly used for P,T metrology in large volume presses. Its precision depends on the contrast between the thermoelastic parameters (bulk modulus $K_0$ and thermal expansion $\alpha$) of the two materials. For example, NaCl and Au are good material choices because NaCl is much more compressible than Au, but has a much lower thermal expansion coefficient. In principle, this method could be applied to a single material with highly anisotropic thermo-elastic properties that could be used as sensor for both P and T variables. As shown in Fig. 9, bP is an excellent candidate material for this type of application. Indeed, as discussed above, it is quasi-incompressible in one direction and has a high linear thermal expansion coefficient ($\alpha_a = 6.46(6)\cdot 10^{-6}$ K$^{-1}$) along the $a$-axis (Fig. 4a, 7a and Table 2). This unique property allows the simultaneous determination of P and T with a good precision.

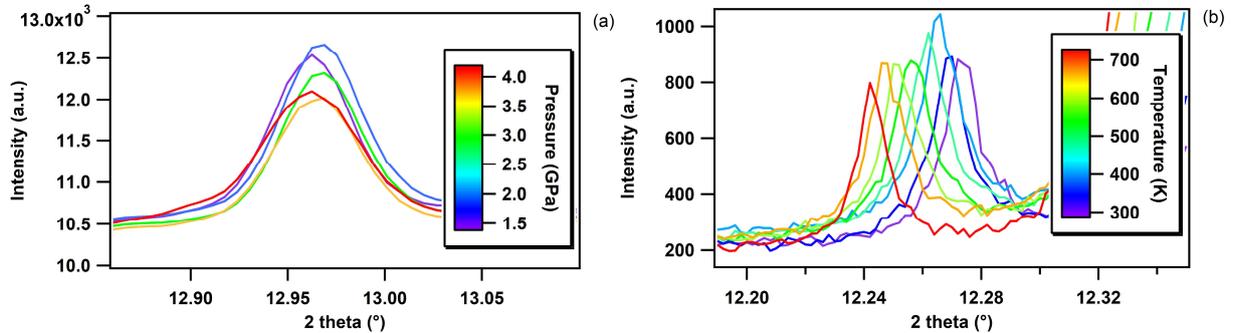

*Fig. 9 (a) Pressure evolution of the position of the (200) reflection of bP at ambient temperature showing no pressure shift due to compression along the a-axis. (b) Temperature evolution of the (200) reflection of bP at ambient pressure.*

Indeed, this remarkable feature makes bP the only known material that can be employed as simultaneous P,T sensor. In practice, its $a$ lattice parameter can be determined, for instance, from the (200) Bragg reflection, the temperature can then be derived using the linear T dependence of $a$ using the equation:

$$T(K) = \frac{a/a_0 - 1}{\alpha_a} + 300 \qquad \text{Eq. 5}$$

Where a and $a_0$ are the values of the a lattice parameter at a given pressure and temperature, and at ambient condition, respectively and $\alpha_a$ is the directional thermal expansion coefficient of bP along *a* (Fig. 4a and Table 2).

The pressure can then be derived from the third-order Birch-Murnaghan equation of state [68] of bP (Eq. 4 and Table 5). The precision of this method is ± 15 K and 0.1 GPa, which is sufficient for the determination of phase diagrams in the P,T range up to 5 GPa and 1700 K.

An example of application which aimed at determining a melting point of bP under pressure using the Paris-Edinburgh (PE) press [57-59] is presented in Fig. 10. The PE press can generate pressures and temperatures in excess of 10 GPa and 2500 K and is ideally suited to determine P, T phase diagrams using *in situ* XRD. As shown in Fig. 10 (left panel), the melting of bP is clearly evidenced by the disappearance of the XRD reflections and appearance of a broad diffuse x-ray scattering signal associated with the loss of the crystalline order. Thanks to the established metrology, the P,T pathway in the phase diagram of bP and position of a melting point have been determined with a precision of ± 15 K at 0.5 GPa. More details about this work will be presented in a future publication.

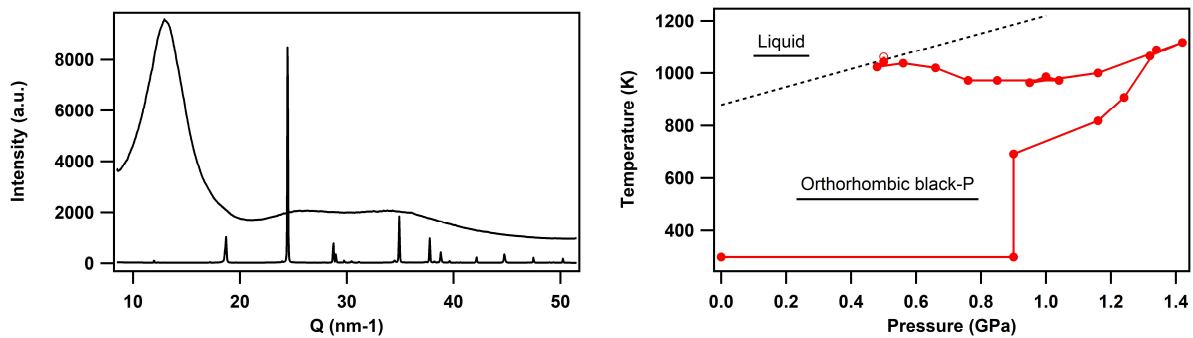

**Fig. 10** *Determination of a melting point of bP using the anisotropic thermo-elastic properties of bP. (Left) Melting criteria: At the melting the Bragg reflections of bP disappear (bottom XRD pattern) and a liquid diffuse signal appears (upper XRD pattern). (Right) P,T pathway of the XRD data collection. Solid red circles indicate crystalline bP, while the empty red circle indicate the liquid state. The dotted line represents the melting line of bP.*

## VI. Conclusions

In this study, we employed *in situ* high-resolution x-ray diffraction to accurately determine the equations of state parameters, as well as the volumetric and linear thermal expansion coefficients, of hexagonal boron nitride (hBN) and black phosphorus (bP). Our investigation focused on the precise characterization of the non-linear variations of the unit-cell parameters of these 2D materials under different temperatures and pressures. In particular, the small in-plane P and T unit cell variations were determined with unparalleled accuracy. These materials possess anisotropic properties due to their layered structure, making them relevant for various energy and technology applications. Our findings offer detailed insights into the structural behavior of hBN and bP at the nanoscale level under high-pressure and temperature conditions achievable with current industrial technology. Consequently, this information contributes to a better understanding and enhancement of the synthesis, stability, and application of these 2D nanostructured materials. Moreover, our data serve as a valuable experimental reference to refine theoretical calculations and gain a deeper understanding of the two-dimensional interactions present in these materials. Lastly, we propose a novel method for high-pressure and high-temperature metrology, utilizing the highly anisotropic directional thermoelastic properties of bP. This innovative approach could be advantageous for large volume press or diamond anvil cell experiments conducted within the low-pressure, high-temperature regime. The exceptional characteristics of bP enable its utilization as a single sensor for simultaneous determination of pressure and temperature through XRD analysis.

## Acknowledgements

We thank the EC through the European Research Council (ERC) for funding the project PHOSFUN ''Phosphorene functionalization: a new platform for advanced multifunctional materials'' (Grant Agreement No. 670173) through an ERC Advanced Grant. We are also grateful to the projects "GreenPhos – alta pressione" (CNR), HP-PHOTOCHEM (Cassa di Risparmio di Firenze) and PRIN 2017 KFY7XF FERMAT "FastElectRon dynamics in novel hybrid-2D MATerials" (MUR). The authors also acknowledge the European Synchrotron Radiation Facility for provision of synchrotron beamtime at the beamline ID27 and the French "Agence Nationale de la Recherche" for financial support under Grant No. ANR-21-CE30-0032-01/03 (LILI).

## Author contributions

The original idea was defined by MM. The experiments were performed by HM, GG, FD, TP, MC, MSR and MM with equal contributions. The data were analysed and the figures produced by HM with contributions from all the co-authors. The manuscript was written by MM, HM, FD and MC with contributions from all the co-authors.